\documentclass{article}

\usepackage{arxiv}

\usepackage[utf8]{inputenc} 
\usepackage[T1]{fontenc}    
\usepackage{hyperref}       
\usepackage{url}            
\usepackage{booktabs}       
\usepackage{amsfonts}       
\usepackage{nicefrac}       
\usepackage{microtype}      
\usepackage{options}
\usepackage{subcaption}
\usepackage{graphicx}
\usepackage{float}
\graphicspath{ {./images/} }

\title{Measuring Mother-Infant Emotions By Audio Sensing}

\author{
  Xuewen Yao \\
  Dpt. of Electrical and Computer Engineering\\
  The University of Texas at Austin\\
  Austin, TX 78712 \\
  \texttt{xuewen@utexas.edu} \\
   \And
  Dong He \\
  Dpt. of Electrical and Computer Engineering\\
  The University of Texas at Austin\\
  Austin, TX 78712 \\
  \texttt{donghe@utexas.edu} \\
  \And
  Tiancheng Jing \\
  Dpt. of Electrical and Computer Engineering\\
  The University of Texas at Austin\\
  Austin, TX 78712 \\
  \texttt{tianchengjing@utexas.edu} \\
  \And
  Kaya de Barbaro \\
  Dpt. of Psychology \\
  The University of Texas at Austin\\
  Austin, TX 78712\\
  \texttt{kaya@austin.utexas.edus} \\
}

\begin{document}
\maketitle
\begin{abstract}
It has been suggested in developmental psychology literature that the communication of affect between mothers and their infants correlates with the socioemotional and cognitive development of infants. In this study, we obtained day-long audio recordings of 10 mother-infant pairs in order to study their affect communication in speech with a focus on mother's speech. In order to build a model for speech emotion detection, we used the Ryerson Audio-Visual Database of Emotional Speech and Song (RAVDESS) and trained a Convolutional Neural Nets model which is able to classify 6 different emotions at 70\% accuracy. We applied our model to mother's speech and found the dominant emotions were angry and sad, which were not true. Based on our own observations, we concluded that emotional speech databases made with the help of actors cannot generalize well to real-life settings, suggesting an active learning or unsupervised approach in the future.

\end{abstract}


\section{Introduction}
Communication of affect between mothers and their infants correlates with infant socioemotional and cognitive development. During interactions between an infant and his/or primary caregiver, the infant responds to the emotional characteristics of the caregiver's behavior in a way that correlates to that emotion. Over time, the infant's emotional reactions are internalized and drive his or her responses to future situations \cite{cohn}. Research in the field has emphasized parent and infant facial expressions for the primacy of facial expressions in theories of emotion, but infants born within six months have relatively poor visual acuity and low contrast sensitivity. In contrast, hearing develops and differentiates earlier than the other senses. Therefore, infants can detect subtle differences in highly similar speech signals \cite{katz}. In addition, the vocal interaction between mother-infant pairs has proved to strengthen attachment and develop the social and linguistic skills of the infant \cite{ferland}. 

Prior work in this subject matter has focused on manually annotating mother-infant interactions, rather than utilizing affective computing for data annotation. Additionally, previous work has been concentrating on the impact of a mother's emotional availability or general interactions on her infant, rather than analyzing the emotions expressed by the mother. 

This study aims to leverage the advancement in affective computing to quantitatively analyze mother-infant interaction with a focus on the mother's speech. We obtained 10 day-long audio recordings of mother-infant interaction in their natural environment and used convolutional neural network (CNN) to build a model for speech emotion classification, a model to apply to  mothers' speech.

Our main contribution in this study is to attempt to leverage affective Computing to quantitatively analyze mother-infant vocal interaction in their natural habitats. Additionally, we used CNN to build a model for speech emotion detection which reached 70\% of accuracy for classifying 6 different emotions.

The remainder of the paper is organized as follows: we first provide a discussion of related works which contain details useful for our study, then we present the implementation details of our work and note the unique contributions our work makes toward identifying emotions portrayed in mother-infant interactions. After that, we discuss lessons learned from this study, future work we'd like to pursue to build on our existing work, and final remarks about our work. Finally, we will note the division of work between each member involved in conducting this study in appendix.

\section{Related Works} \label{related}
Numerous psychology studies have emphasized the significance of emotional availability in a mother-infant relationship. Field, Vega-Lahr, Scafidi, and Goldstein suggest that the emotional unavailability of a mother is more troubling to an infant than her physical absence \cite{field}. Sorce and Emde also observe the negative impact a mother's emotional unavailability has on her infant, and additionally indicate that the emotional availability of a mother yields significant positive effects on her child's affective, social, and exploratory behaviors \cite{sorce}. The emotional availability a mother displays to her child also exposes the child to a wide range of emotions. Trainor observes that emotional expression is very similar in infant-directed and adult-directed speech, and there is a widespread expression of emotions to infants in comparison to those in typical adult interactions \cite{trainor}. These works are novel in demonstrating the significant impact a mother has on her infant when interacting with him or her verbally, which is a key motivation for our study.

However, these studies focus on the impact of the characteristics that define a mothers' emotional availability, such as facial expression, rather than the emotions expressed by the mothers when interacting with their infants. Additionally, these studies do not utilize computational methods to analyze the interactions. In contrast, our study focuses solely on computational audio sensing of mother-infant interactions with a focus on detecting the emotions portrayed in the mother's voice.

In our study, LENA was used to record and annotate audio data. LENA is the industry standard for measuring vocals with children up to three years in age. LENA consists of a device and software. The LENA device is a single integrated microphone that records unprocessed acoustic data. The LENA software contains algorithms that break up the acoustic signal into segments that are labeled by the primary sound source \cite{LENA}.

Advancements in computer science has allowed researchers to explore effective ways to detect emotions through audio sensing. Rachuri et al. introduced a mobile sensing platform called EmotionSense that gathers a user's emotions as well as proximity and patterns of conversation by processing outputs from built-in sensor in off-the-shelf smartphones. The emotion recognition component is built by a Gaussian Mixture Model that is trained using an existing emotional speech database. Anger, fear, neutral, sadness, and happiness were the emotion classes used for training the data. This subsystem produced 71\% accuracy when performing emotion detection \cite{rachuri}. 

Lane et al. build an audio sensing framework, called DeepEar, that utilizes four coupled Deep Neural Networks (DNNs). The DNN layer that is particularly interesting given the context of our study is the emotion recognition DNN. Like the other DNNs, this layer is trained with labeled data from the same emotional speech database that was used in the EmotionSense study. The emotion classes utilized were anger, fear, neutral, sadness, and happiness. This layer was also trained with unlabeled data gathered from various environments. This method resulted in about an 80\% accuracy for emotion detection \cite{lane}.  These previous works demonstrate promising approaches to detecting emotion with audio sensing, especially using deep learning and they were taken into consideration when we built a model for detecting mother emotions in our study.

\section{Method}
\subsection{Dataset}
Upon IRB approval, we obtained a total of 10 different audio recordings from Daily Activity Lab. These recordings came from day-long interactions of 10 mother-infant pairs. LENA device attached to infant's vest was used for data collection. The mean data length of audio recordings were 16.1 hours (std: 4.6; range: 3.22 - 18.63 hours).

\begin{figure}[H]
\centering
\includegraphics[width=0.6\columnwidth]{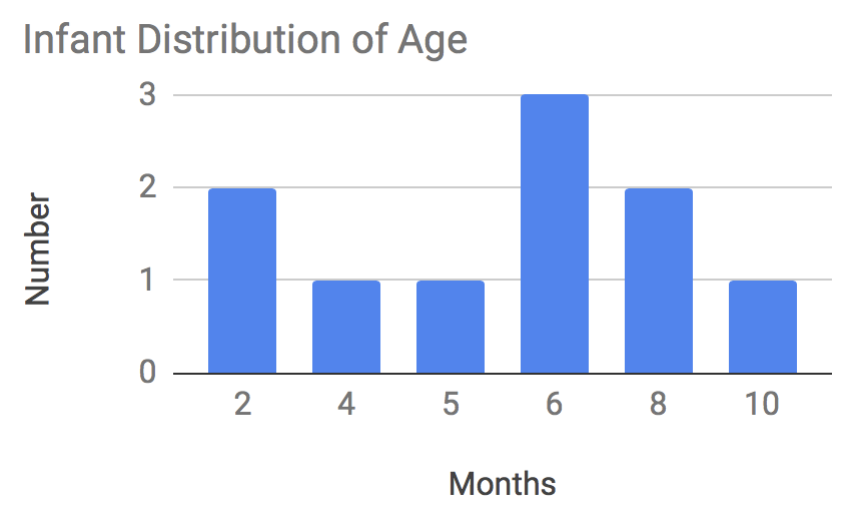}
\caption{Infant Distribution of Age}
\label{age}
\end{figure}

In terms of mothers and their infants, the mean age of infants were 5.74 months (std: 2.51; range: 1.7 - 10.4 months) and there were 6 female, 4 male in the dataset. Detailed infant age distribution can be found in Figure \ref{age}. The mean age of mothers were 32.33 years old (std: 4.55; range: 29 - 41 years). Of 10 mothers, 7 of them were White, 1 was Hispanic and 2 were mixed. 

\subsection{Preprocessing}
In order to extract mother's speech from all recordings, we considered the pipeline of speech activity detection, speaker diarization and speaker identification. This approach was abandoned because we could not reach an high accuracy during speaker diarization. Instead, we turned to LENA software.

As mentioned in Section \ref{related}, LENA software can break up the acoustic signals into segments and label each segment based on primary sound source.  These sound sources include Female Adult Near (FAN), Female Adult Far (FAF), Male Adult Near (MAN), Male Adult Far (MAF), Key Child Clear (CHN) and so on. We decided to use signals labeled FAN as an approximation of mother's speech. This approximation was reasonable but not entirely accurate for it could include the voices from other female family members and friends, but it was the best option for us. We did not use FAF as the loudness of speech could affect emotion detection. 

After using LENA software to extract female's speech, a total of 15411 files were obtained from 10 sessions. Each session had an average of 1541 files (std: 523). We randomly selected 100 audio files and confirmed that 90 of them were female's speech.

\subsection{Model}
With the large number of audio files segmented using LENA and lack of trained ears, it was hard for us to manually label the dominant emotions within each file.

Thus, in order to build a reliable speech emotion detection model, we had to use other databases. We chose RAVDESS \cite{ravdess} which was publicly available online. The database contained 7356 files (total size: 24.8 GB; including both audios and videos) with 24 professional actors (12 female, 12 male) vocalizing in a neutral North American accent. In this study, we used 1012 female audio files containing 6 emotions: neutral, calm, happy, sad, angry, fearful.	

We extracted 5 acoustic features from the raw audio data. These features are MFCC, MFCC delta, MFCC delta-delta, Zero-Crossing Rate and root-mean-square (RMS) energy. MFCC, MFCC delta and MFCC delta-delta represent the spectral content of audio signals and MFCC was mentioned in \cite{tal} as one of the vocal features of emotions. Zero-Crossing Rate was to extract pitch and RMS energy loudness of speech signals.

\begin{figure}[H]
\includegraphics[width=1.0\columnwidth]{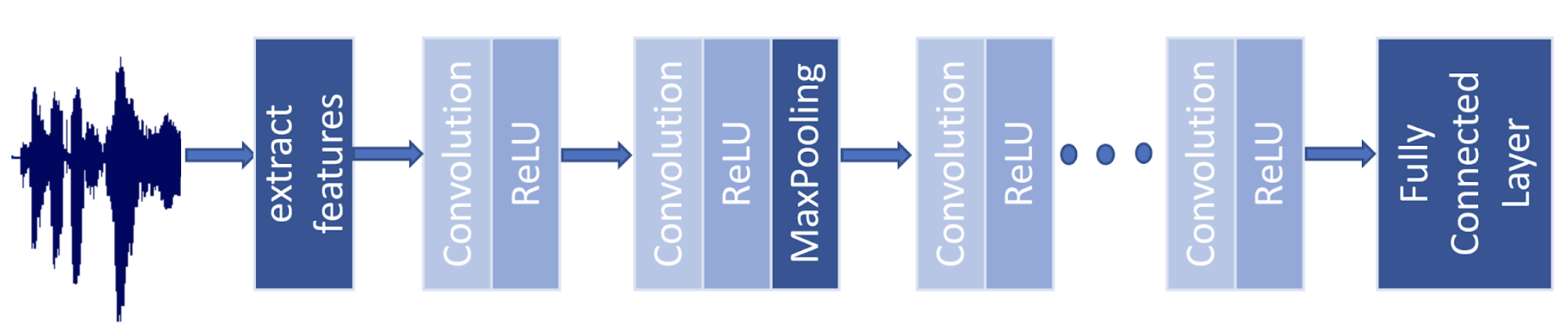}
\caption{Model}
\label{model}
\end{figure}

These features were fed into a Convolutional Neural Net for training. Features instead of raw data were used as input because of our lack of computing power. The structure of our neural nets can be found in Figure \ref{model}. Our model contained 6 convolutional layers, 6 RELU activation layers, 1 pooling layer and 1 fully connected layer. The batch size was 25 and RMSProp optimizer was used with learning rate 0.0001.

After around 300 epochs (see Figure \ref{accuracy}), the training accuracy reached 99\% while testing 70\%.
The confusion matrix is shown in Figure \ref{cm}. With this reliability obtained, we applied the model to each of the mother utterances extracted using LENA. The results are shown in Section \ref{results}.

\begin{figure}[H]
    \centering
    \begin{subfigure}[b]{0.45\textwidth}
        \includegraphics[width=\textwidth]{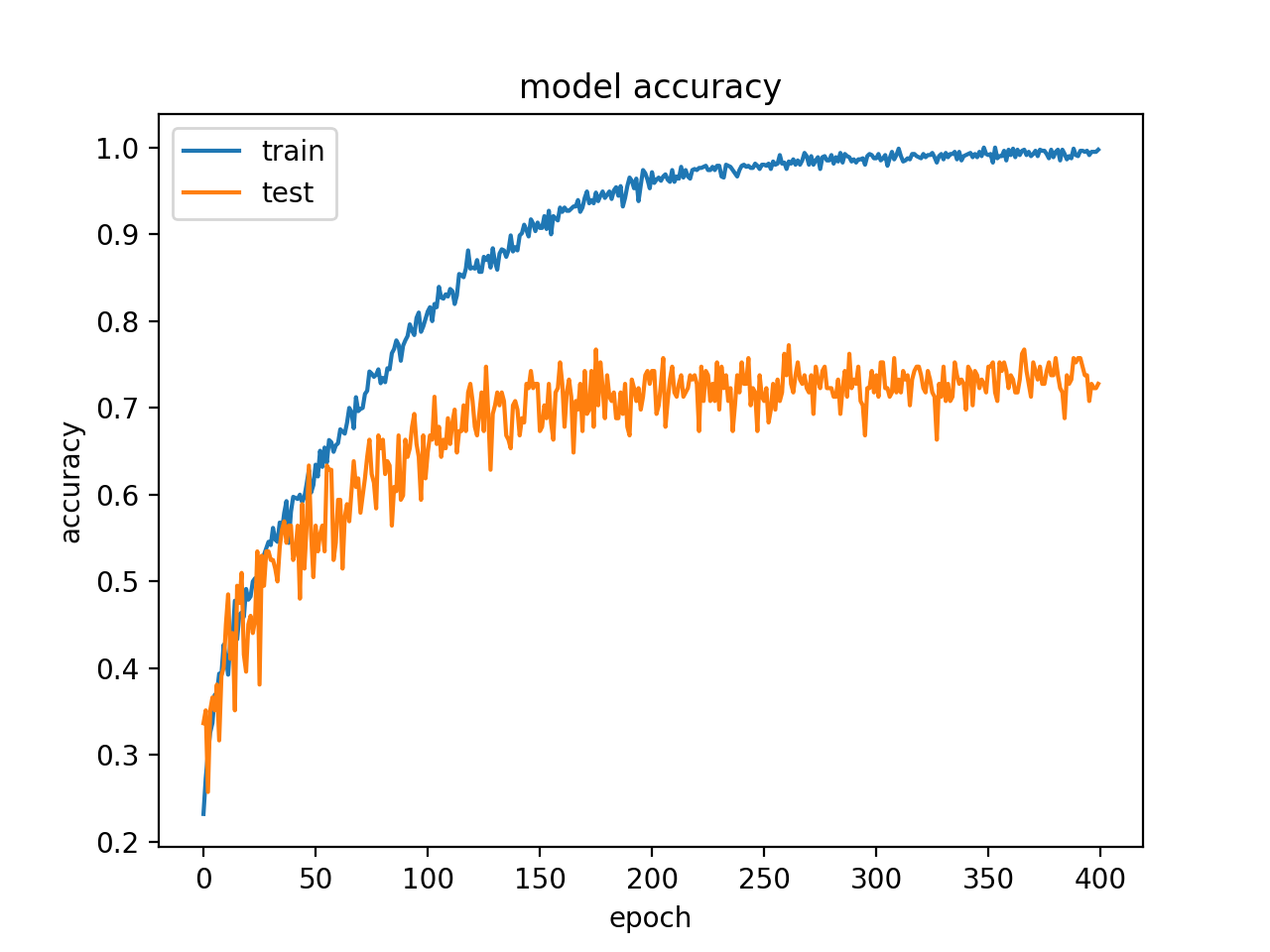}
        \caption{Training and testing accuracy for our CNN models. Both accuracy converged.}
        \label{accuracy}
    \end{subfigure}
    \begin{subfigure}[b]{0.45\textwidth}
        \includegraphics[width=\textwidth]{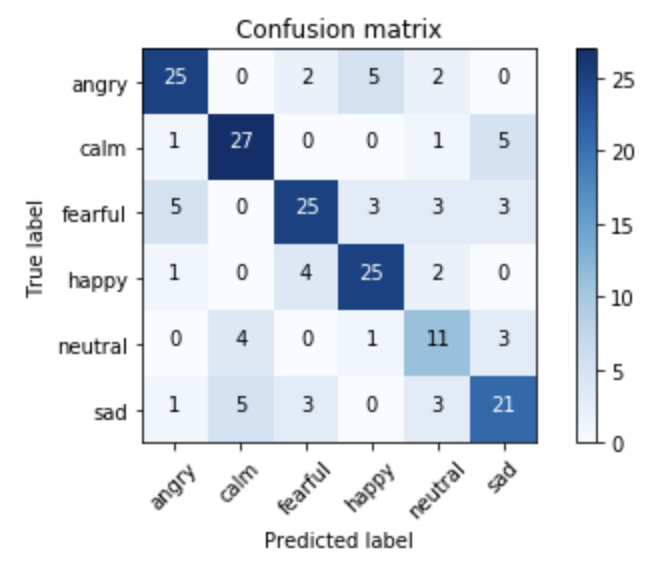}
        \caption{Confusion Matrix for RAVDESS test dataset. It shows our model's ability to predict accurately for each class.}
        \label{cm}
    \end{subfigure}
\end{figure}

\section{Results} \label{results}
\begin{figure}[H]
\centering
\includegraphics[width=0.6\columnwidth]{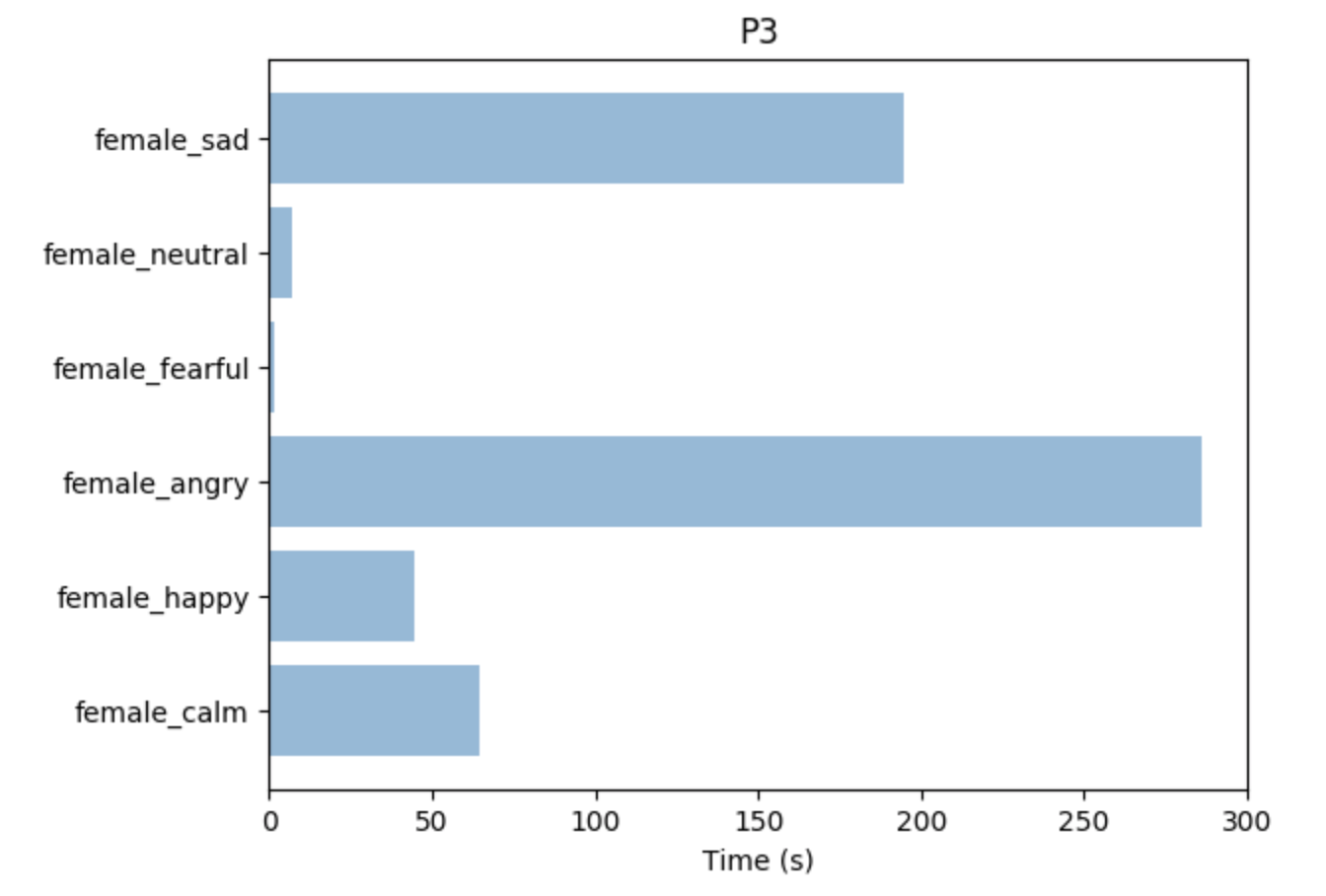}
\caption{Result of P3 is shown as an example here. The dominant emotions were angry and sad, which was wrong.}
\label{example_result}
\end{figure}

Figure \ref{example_result} shows the result of P3 as an example. It can be seen that the dominant emotions were angry and sad, followed by calm and happy. And there was little neural and fearful in the voice.

This result differed from our understanding and observations and it populated in all sessions with little variation. The results for all mother-infant pairs can be found in Appendix \ref{app} and we will discuss the implications of results in Section \ref{implications}.

\section{Discussions}
\subsection{Implications} \label{implications}
The results given by our model differed from our own understanding and it propelled us to think the reason behind it. 

After listening to hundreds of female speech recordings, we observed that mother's speech was usually happy and neutral. However, comparing to the speech used in RAVDESS database, mother's speech in real-life settings was much faster and sometimes elevated. In addition, there were background noises, such as infant's fussing in the recordings. All these factors could negatively affect the classification results. Thus, here we question the suitability of building and later testing the model using completely different datasets and whether models built using lab data can generalize to in-the-wild data. We do believe it is a common concern for many other researchers who want to test the model (built in lab) in the wild. This concern can lead to other approaches such as active learning or unsupervised learning.

\subsection{Future Work}
In our study we focused on detecting emotions of mothers in the presence of their infants using audio sensing. In the future, if we obtain a reliable model for real-life settings, we'd like to use affective computing for detecting emotions expressed vocally by infants during interactions with his or her primary caregiver. By doing this, we hope to provide a unique insight into the impact of a caregiver's emotional expressions on an infant. 

In addition to analyzing infant speech, we would like to extend our model to classify emotions in the father's speech during father-infant interactions. Not only would this make our model more generalized, but it would also be useful in observing the emotions portrayed in situations where mothers and fathers interact with their infant together.

\section{Conclusion}
In this study, we extracted and analyzed female voices present in natural interactions between mothers and their infants recorded during day-long natural interactions. We built a deep learning model using CNNs and online database (RAVDESS) to classify emotions present in the female voices. Using this model, we obtained 70\% accuracy in classifying six different emotions: neutral, calm, happy, sad, angry, and fearful. When applying our deep learning model to mother's speech, we found that the dominant emotions were angry and sad, which was contrary to our observations, suggesting future approach using active learning or unsupervised learning. This study attempts and utilizes affective computing to provide valuable insight into the emotions projected from a mother's speech while interacting with her infant. Given the results of this study, our experience could be useful for future works in emotion detection and developmental psychology.

\appendix
\section{Results for all mother-infant pairs} \label{app}

\begin{figure}[H]
    \centering
    \begin{subfigure}[b]{0.48\textwidth}
        \includegraphics[width=\textwidth]{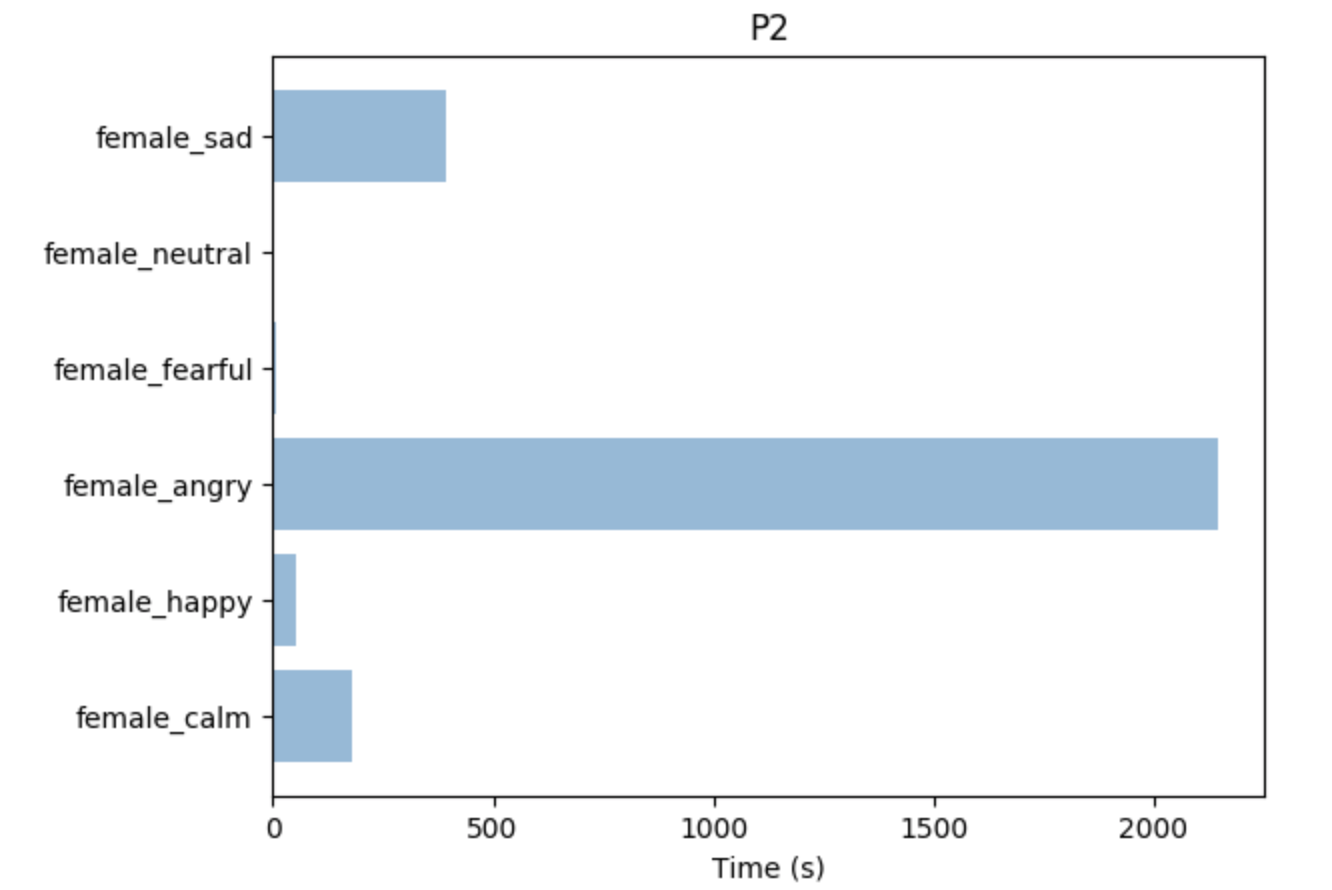}
        \caption{P2}
        \label{accuracy}
    \end{subfigure}
    \begin{subfigure}[b]{0.48\textwidth}
        \includegraphics[width=\textwidth]{P3}
        \caption{P3}
        \label{cm}
    \end{subfigure}
\end{figure}

\begin{figure}[H]
    \centering
    \begin{subfigure}[b]{0.48\textwidth}
        \includegraphics[width=\textwidth]{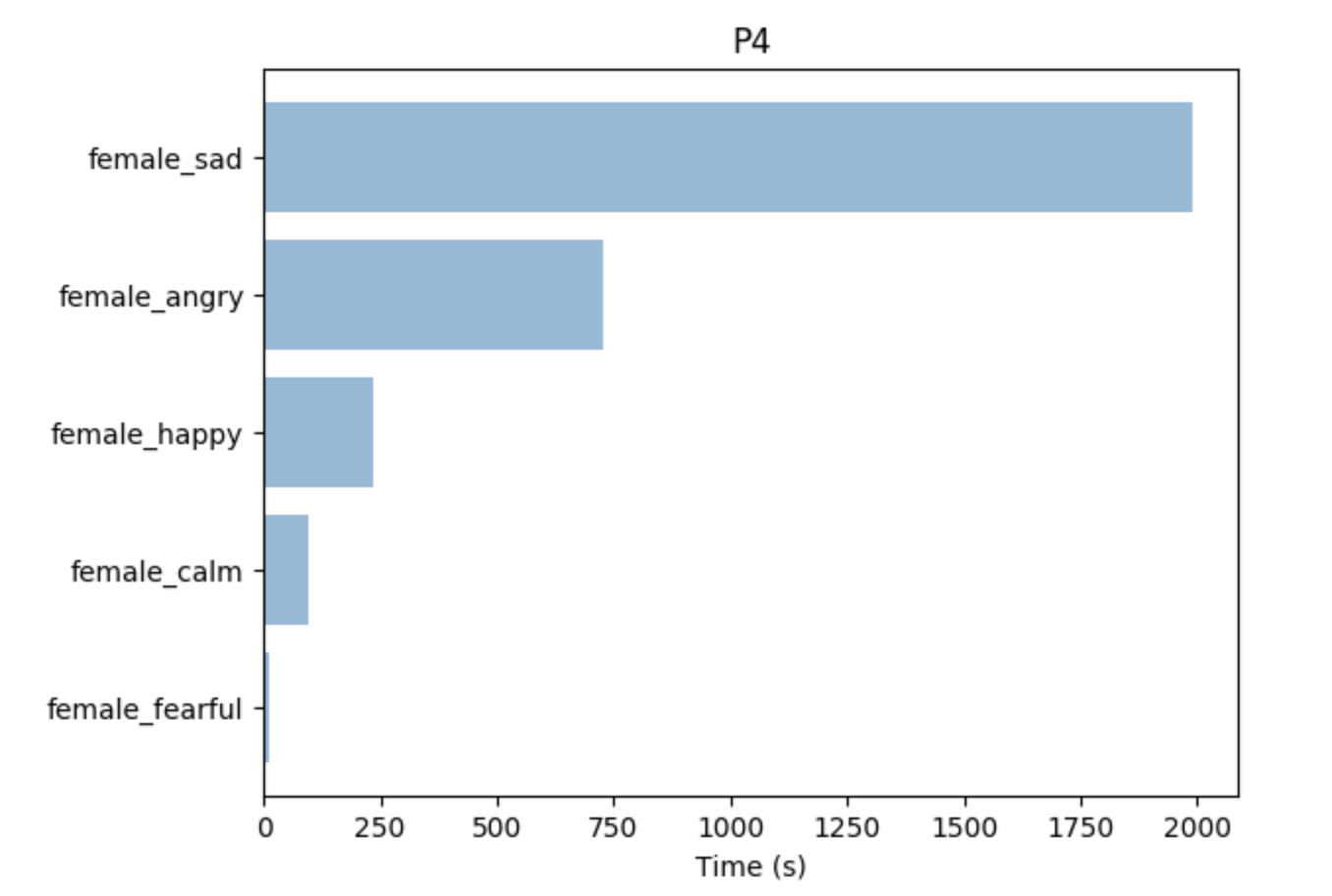}
        \caption{P4}
        \label{accuracy}
    \end{subfigure}
    \begin{subfigure}[b]{0.48\textwidth}
        \includegraphics[width=\textwidth]{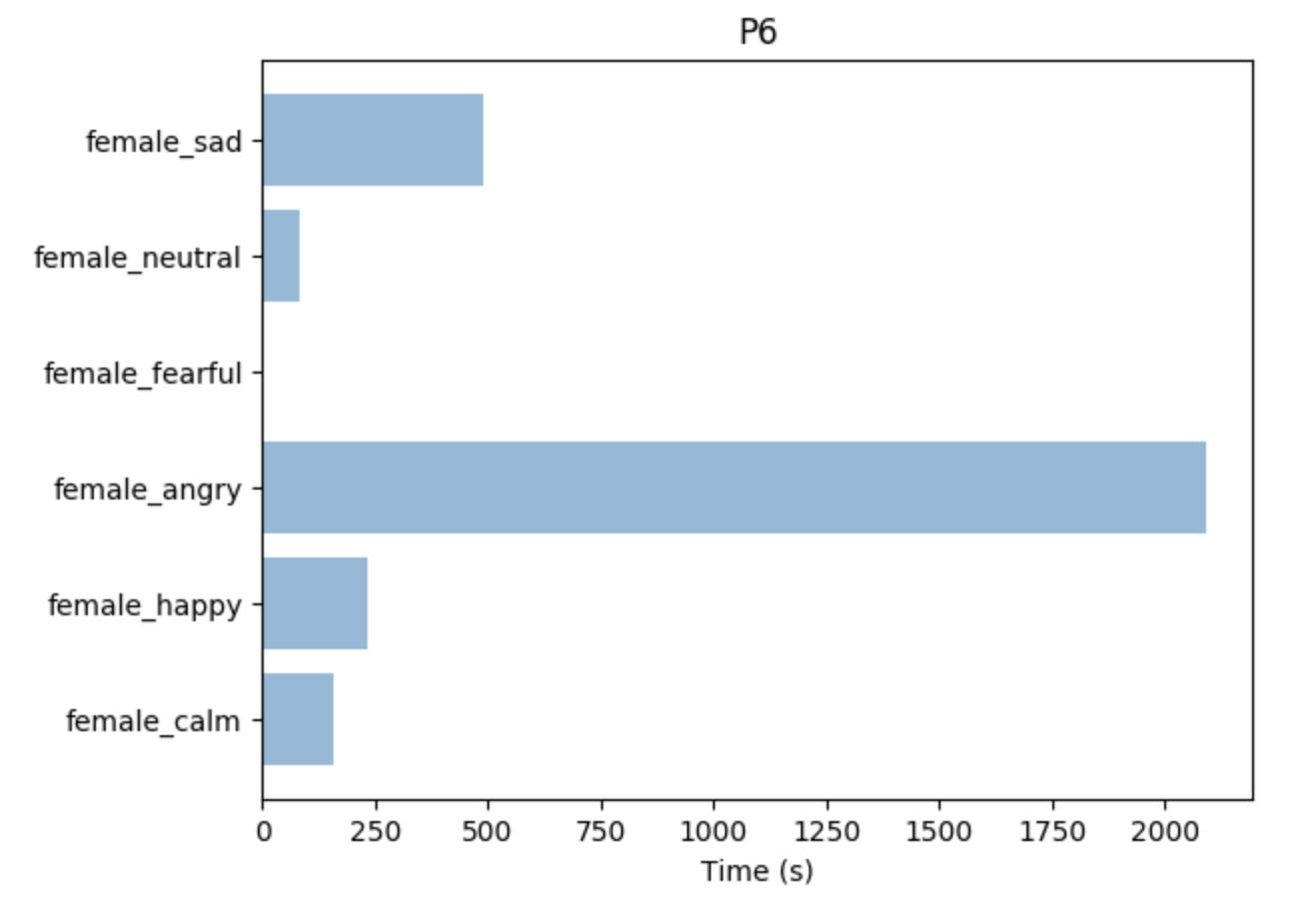}
        \caption{P6}
        \label{cm}
    \end{subfigure}
\end{figure}

\begin{figure}[H]
    \centering
    \begin{subfigure}[b]{0.48\textwidth}
        \includegraphics[width=\textwidth]{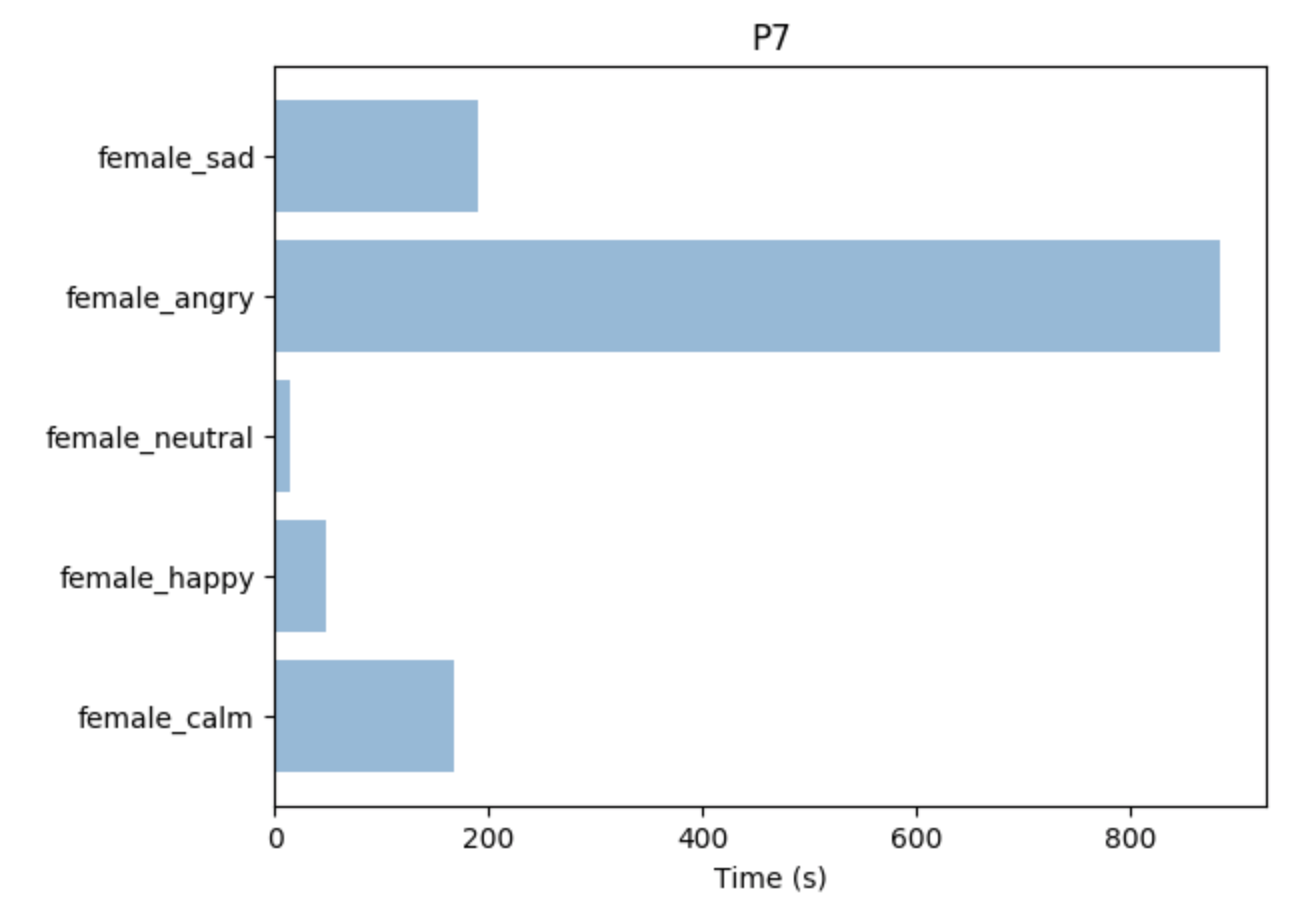}
        \caption{P7}
        \label{accuracy}
    \end{subfigure}
    \begin{subfigure}[b]{0.48\textwidth}
        \includegraphics[width=\textwidth]{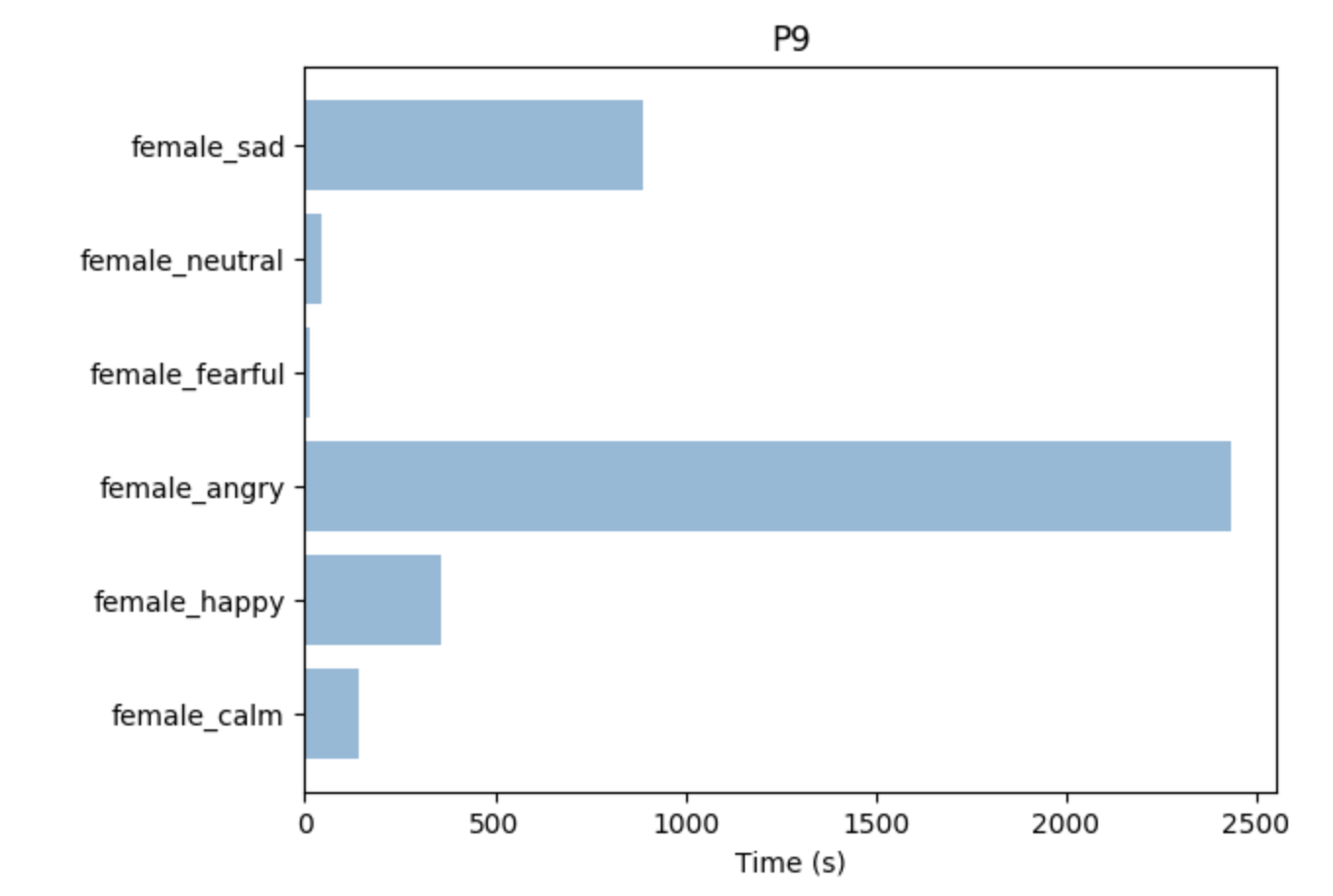}
        \caption{P9}
        \label{cm}
    \end{subfigure}
\end{figure}

\begin{figure}[H]
    \centering
    \begin{subfigure}[b]{0.48\textwidth}
        \includegraphics[width=\textwidth]{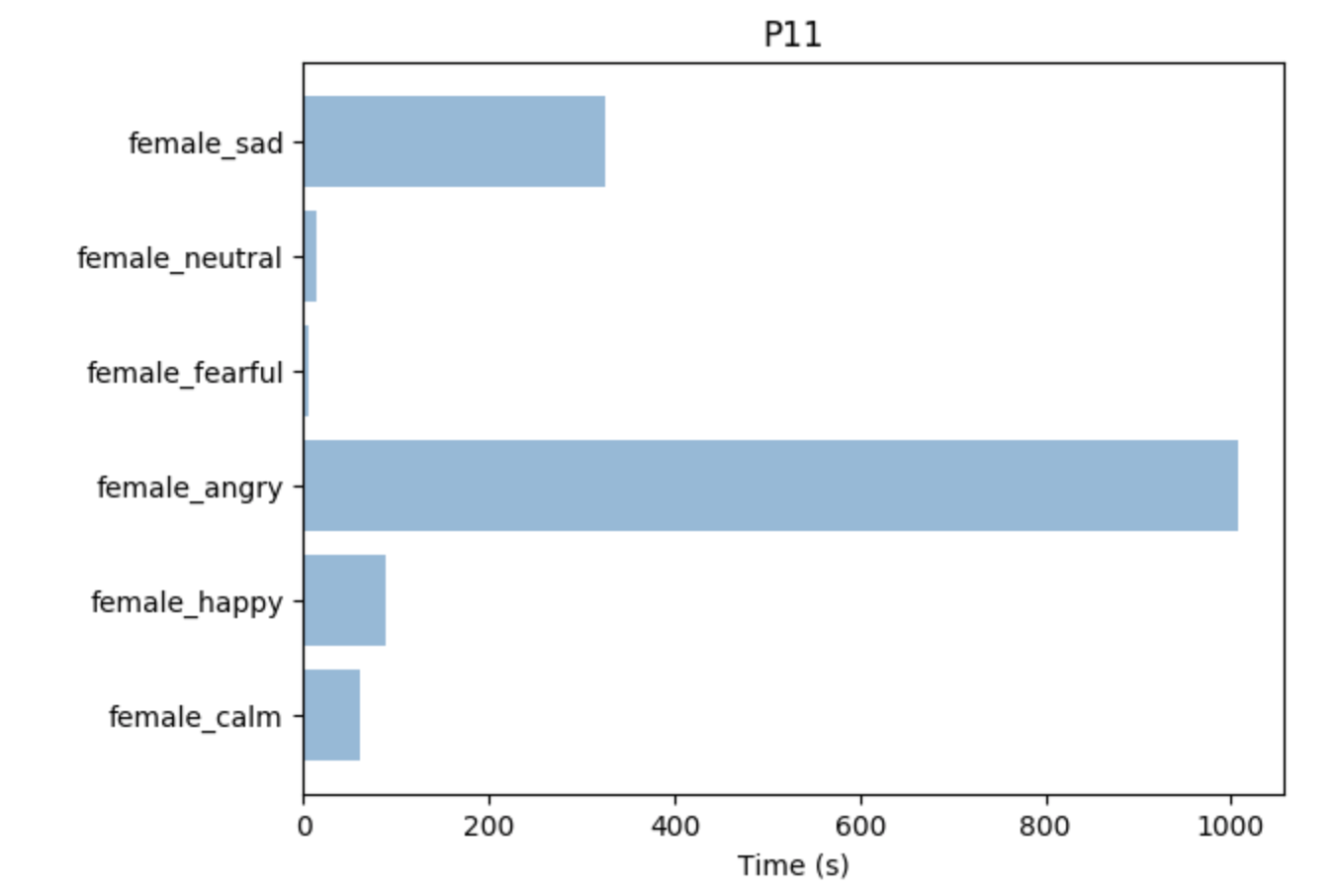}
        \caption{P11}
        \label{accuracy}
    \end{subfigure}
    \begin{subfigure}[b]{0.48\textwidth}
        \includegraphics[width=\textwidth]{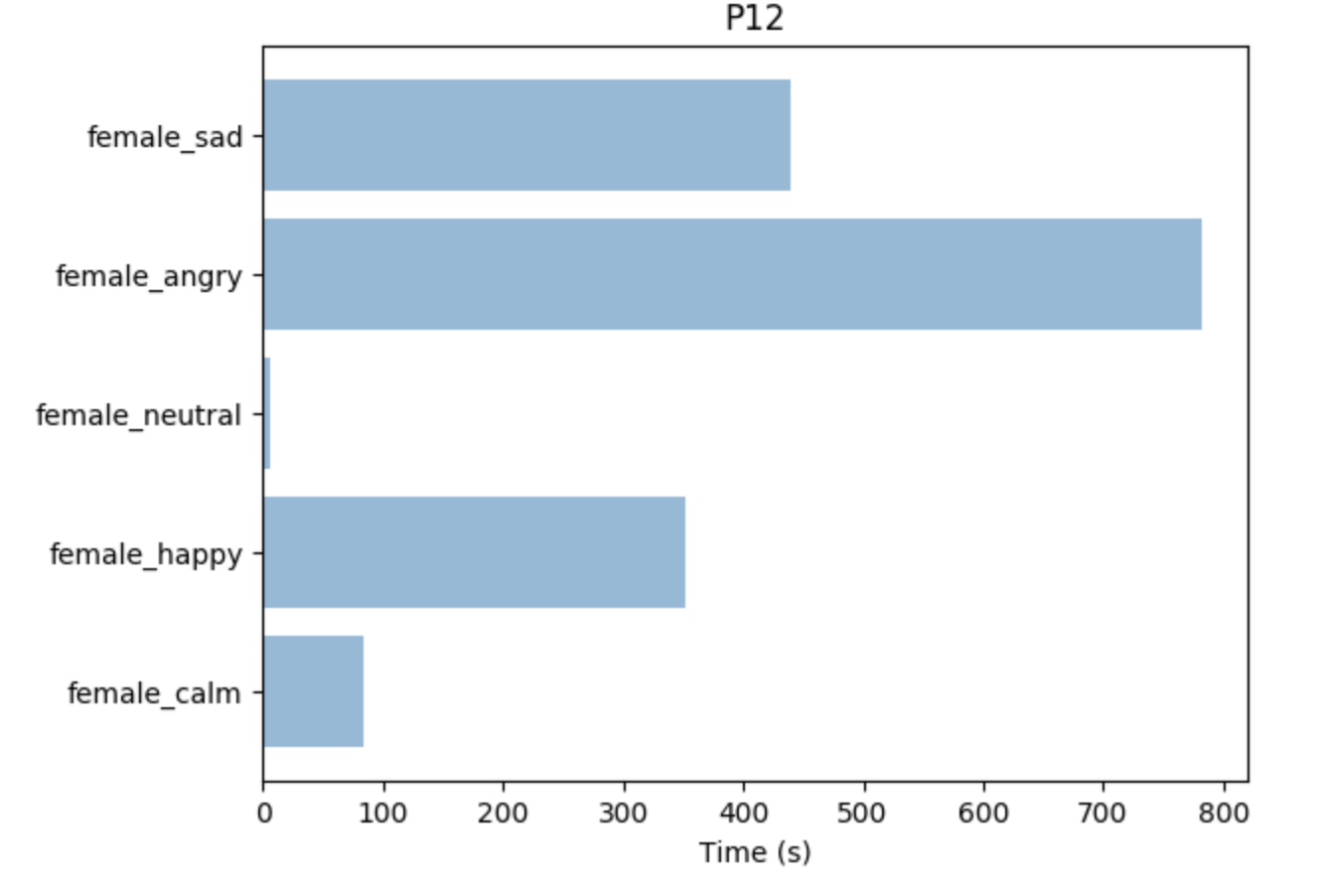}
        \caption{P12}
        \label{cm}
    \end{subfigure}
\end{figure}

\begin{figure}[H]
    \centering
    \begin{subfigure}[b]{0.48\textwidth}
        \includegraphics[width=\textwidth]{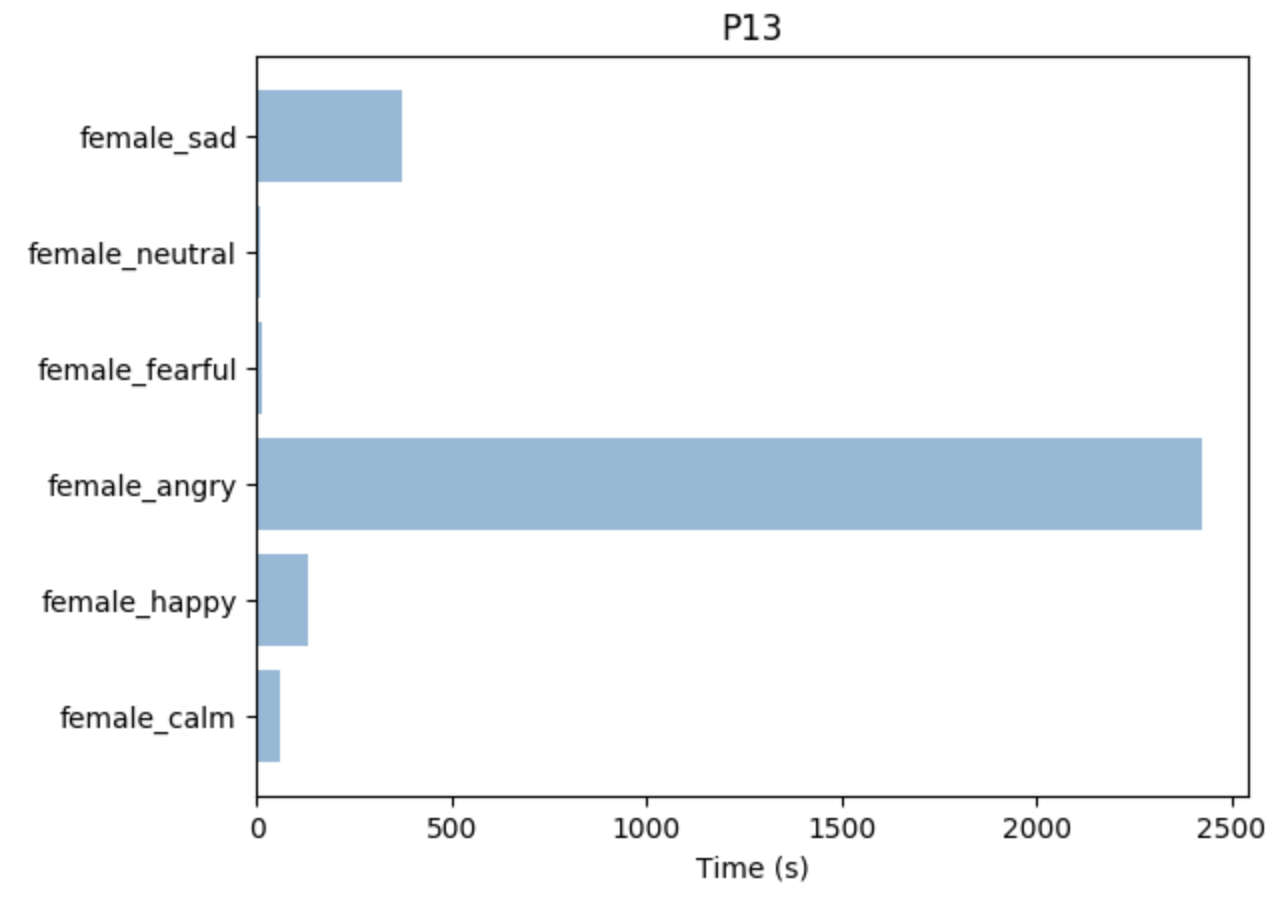}
        \caption{P13}
        \label{accuracy}
    \end{subfigure}
    \begin{subfigure}[b]{0.48\textwidth}
        \includegraphics[width=\textwidth]{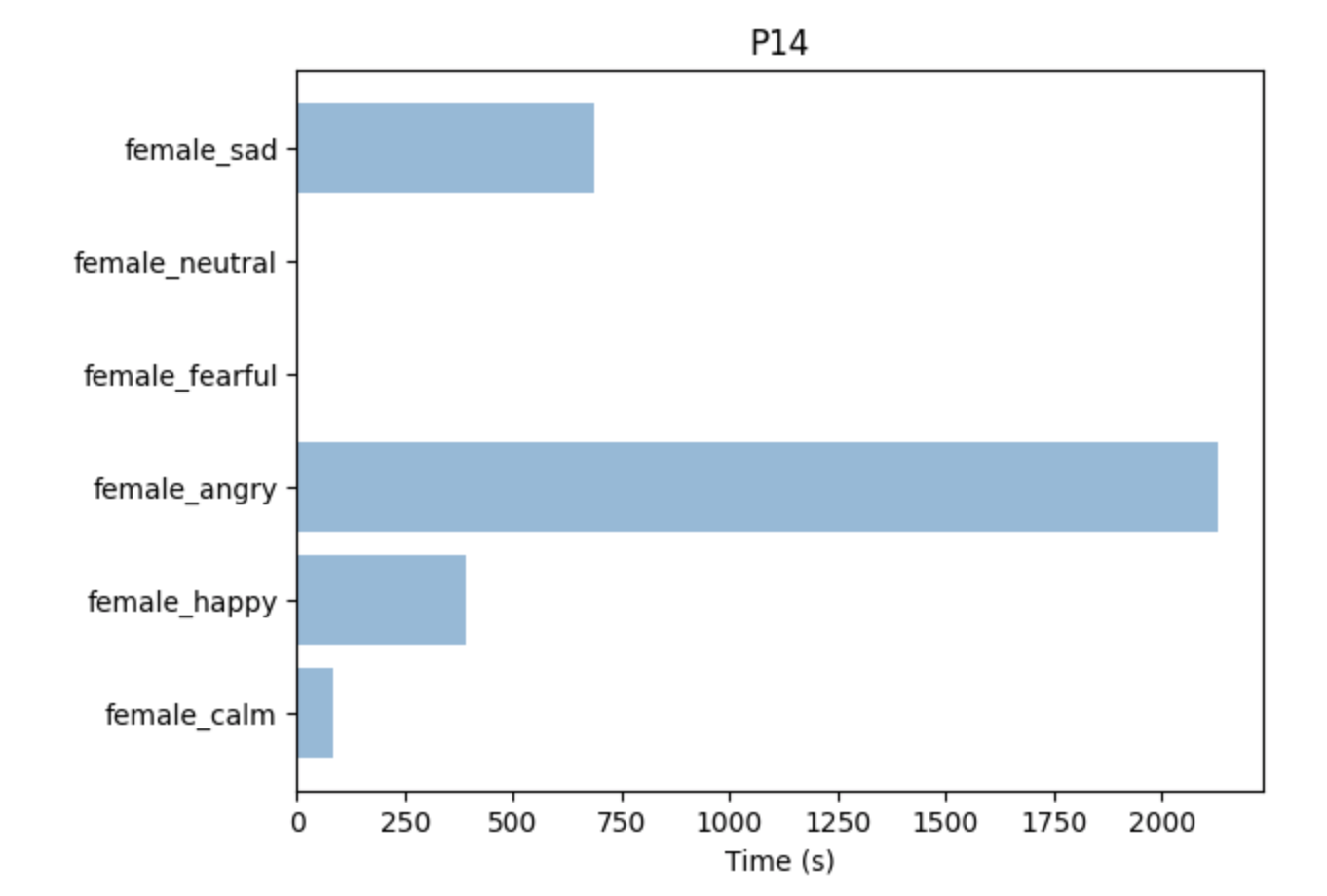}
        \caption{P14}
        \label{cm}
    \end{subfigure}
\end{figure}

\end{document}